\documentclass{aa}  

\usepackage{graphicx}
\usepackage{enumerate, graphicx}
\usepackage{float}
\usepackage{txfonts}
\usepackage{xcolor}

\begin{document} 

   \title{The VVVX quest for satellites
around the Circinus galaxy}
   \titlerunning{Searching for Circinus satellites}

   \author{L. D. Baravalle
          \inst{1,2}
          \and A. L. O'Mill\inst{1,2}  \and  M. V. Alonso\inst{1,2} \and C. Obasi\inst{3} \and D. Minniti\inst{3,4} \and 
          M. G\'omez\inst{3} \and  C. Villalon\inst{1,2} \and J. Nilo-Castell\'on\inst{6}  \and C. Valotto\inst{1,2} \and M. Soto\inst{7} \and I. V. Daza Perilla\inst{8,9} \and M. A. Sgr\'o\inst{9,2}  \and  J. G. Fernández-Trincado\inst{10}} 
          
         \institute{Instituto de Astronomía Teórica y Experimental, (IATE-CONICET), Laprida 854, X5000BGR, Córdoba, Argentina.
         \and Observatorio Astronómico de Córdoba, Universidad Nacional de Córdoba, Laprida 854, X5000BGR, Córdoba, Argentina.
         \and  Instituto de Astrofísica, Dep. de Física y Astronomía, Facultad de Ciencias Exactas, Universidad Andres Bello, Av. Fernández Concha 700, Santiago, Chile.
         \and Vatican Observatory, V00120 Vatican City State, Italy.
         \and Departamento Astronom\'ia, Universidad de La Serena. Av. Raúl Bitran 1305, La Serena, Región de Coquimbo, Chile. 
          \and Instituto de Investigación en Astronomía y Ciencias Planetarias, Universidad de Atacama, Copayapu 485, Copiapó, Chile 
         \and Center for Space Science and Technology, University of Maryland, Baltimore County, 1000 Hilltop Circle, Baltimore MD 21250, USA.
        \and
        Center for Research and Exploration in Space Science and Technology, NASA/Goddard Space Flight Center, Greenbelt, MD 20771, USA. 
        \and  Instituto de Altos Estudios Espaciales “Mario Gulich” (CONAE – UNC), Argentina.
         \and Instituto de Astronomía, Universidad Católica del Norte, Av. Angamos 0610, Antofagasta, Chile.
         }


   \date{Received XXX; accepted XXX}

 
  \abstract
   {The Circinus galaxy is the nearest type-2 Seyfert galaxy, which is at a distance of 4.2 Mpc. Its environment is challenging to explore because this galaxy is located at low Galactic latitudes, behind the disc of the Milky Way.} 
   {The long-term goal is to characterise the Circinus galaxy halo,
   and we are interested in determining the possible presence of dwarf satellites using near-infrared data.
   }
   {We selected 1,542 galaxies from the VVV near-IR galaxy catalogue within a 2-degree radius around the Circinus galaxy, representing roughly 2/3 of the virial radius. Structural parameters such as half-light radii and colours were used, and correlations were examined. A neural network was trained with 486 galaxies with known spectroscopic redshifts in order to estimate photometric redshifts for all of these galaxies. Potential satellites of the Circinus galaxy were defined based on half-light radii compatible with the typical sizes of dwarf satellites in galaxies, and combined with photometric redshifts.
   }
   {The galaxy properties are reliably characterised down to $K_{s} \sim 15.5$ mag, which represents about 90\% completeness of the detections. 
   At the distance of the Circinus galaxy, this limiting magnitude corresponds to $K_{s}$ absolute magnitude of $-12.6$ mag, which allows us to find dwarf galaxies. 
   {There are 20 galaxies with half-light radii larger than 2.45 arcsec, however,}  only 8 of these have photometric redshifts below 0.04. None of these galaxies is close to the Circinus galaxy, which has a redshift of 0.0015, thus showing no evidence of associated clustering. The ANNz model exhibited a high degree of accuracy in the range of $0.001 < z_{phot} < 0.023$, therefore validating this method in these extinct regions.  }
   {The presence of dwarf satellites associated with the Circinus galaxy could not be confirmed with the available data in the studied region. The apparent lack of satellites may be genuine, possibly related to AGN feedback effects.  
   Nevertheless, this work demonstrates the effectiveness of combining near-infrared data and machine learning techniques to estimate photometric redshifts 
   at low Galactic latitudes, thus providing useful information for
   future spectroscopic follow-up campaigns. }


   \maketitle

\section{Introduction}

The Circinus galaxy (ESO 97-G13) is a nearby spiral galaxy located at J2000 coordinates RA = 14h 13m 09s.906, Dec = $-65^{\circ} 20^{\prime} 20.47^{\prime\prime}$ ($l = 311.326^{\circ}$, $b = -3.808^{\circ}$). It is among the nearest active galaxies to the Milky Way (MW), at a distance of $D = 4.2 $ Mpc \citep{Tully2009}. 
The Circinus galaxy is classified as SA(s)b, being an unbarred spiral galaxy with moderately wound arms and hosting a type-2 Seyfert nucleus which exhibits narrow emission lines and an obscured active galactic nucleus (AGN, \citealt{Oliva1994,Maiolino1998,Stalevski2017}).  
The systemic velocity of the galaxy is 439 km/s based on HI measurements reported by \cite{Jones1999}, corresponding to a redshift of 0.0015. The total stellar mass of the galaxy was estimated by \cite{For2012} to be 9.5 $\times$ 10$^{10}$ \(\textup{M}_\odot\). Using the stellar mass-halo mass relation \citep{Behroozi2010}, the Circinus halo mass is approximately 1 $\times$ 10$^{13}$ 
\(\textup{M}_\odot\).
The logarithm of $K_s$ luminosity of the galaxy in solar units, corrected for extinction, is 10.60 \citep{Karachentsev2013}, with the virial radius being 220 kpc \citep{Mutlu2021}.

The study of the Circinus galaxy presents considerable challenges due to its location at low latitudes behind the Galactic disc, where the effects of interstellar extinction and crowding are particularly pronounced. The majority of previous research has been focused on its active galactic nucleus, super-massive black hole \citep{smith2001chandra}, and X-ray emission \citep{andonie2022multiwavelength}, with a few studies conducted on optical passbands.
The distribution of gas in the Circinus galaxy exhibits an asymmetric pattern, which might be indicative of galaxy interactions \citep{Jones1999}.
According to the nearby galaxy catalogue \citep{Karachentsev2013}, there are 32 galaxies within a velocity range of $\pm$ 200 km/s relative to the Circinus galaxy, with the closest of these galaxies being located at an angular separation of 16 degrees, corresponding to a distance of approximately 1.1 Mpc. The galaxy HI-ZOA J1353-58 was identified by \cite{Staveley2016} as being a potential companion of the Circinus galaxy. It is less than 8 degrees, or about 3 virial radii, from the Circinus galaxy, and has a recessional velocity difference of about 200 km s$^{-1}$. 

In \cite{Bullock2017} review on challenges at small scales to the standard cosmological model, known as the $\Lambda$ Cold Dark Matter paradigm, one of the most prominent issues discussed is the problem of missing satellites \citep{Klypin1999}. Understanding the distribution of satellites in galaxies is therefore a key factor in galaxy formation models. Identification of satellite galaxies beyond the Local Group represents a significant challenge, even in relatively nearby galaxies with distances smaller than 4
Mpc \citep[e.g.,][]{Webster1979,Rejkuba2007,Crnojevic2014,Crnojevic2016, MartinezDelgado2021,MutluPakdil2022,MutluPakdil2024,Doliva-Dolinsky2025}. Using simulated galaxies to be observed at the  Rubin Observatory, \cite{Mutlu2021} studied dwarf detection efficiencies for a variety of distances, luminosities, and sizes, and identified regions in the luminosity-size space where the galaxy satellites should be located.  Galaxies with similar stellar masses as the Circinus galaxy 
show a wide dispersion in the number of confirmed satellites. For instance, the MW and Andromeda host 49 and 39 satellites, respectively \citep{Smercina2018}, while M94 has only two within 150 kpc. This scatter is not unexpected as it might reflect both the intrinsic stochasticity in halo assembly histories and the varying impact of baryonic feedback processes, such as AGN-driven outflows, which can suppress the formation of faint satellites. 

The VISTA Variables in the V\'ia L\'actea survey (VVV, \citealt{Minniti2010}) and its extension (VVVX, \citealt{Saito2024}) 
were designed to map the Galactic bulge and disc in $ZYJHK_s$ and $JHK_s$ near-infrared (NIR) passbands, respectively. These surveys reached 5$\sigma$ depths of $K_s$  $\sim$ 17.5 – 18.0 magnitudes, covering 560 square degrees for the VVV survey and 1200 square degrees for the VVVX survey.
Data were obtained using the Visible and Infrared Survey Telescope for Astronomy (VISTA; \citealt{Emerson2006}) 4.1 m telescope at the Paranal Observatory of the European Southern Observatory (ESO) in Chile, equipped with the VISTA InfraRed CAMera (VIRCAM). The main goal of both surveys was to monitor the variability of different types of stars and to determine the three-dimensional structure of the MW. 
The NIR observations also mitigate the issues of interstellar extinction and stellar crowding, thereby offering a distinctive opportunity for detecting and studying galaxies situated beyond the Galactic plane.  

\cite{Obasi2023,Obasi2024} searched and characterised the Circinus globular cluster (GC) system. NIR and optical imaging, in conjunction with data from the Gaia mission \citep{Gaia2016}, have been used to reveal a population of 111 GC candidates. 
However, perhaps the most striking finding was that the Circinus galaxy is potentially the host of a large GC system that may extend up to several tens of kpc. A few galaxies have been identified in the Circinus surroundings, but so far no known satellites have been confirmed. 
In the Circinus region, there is considerable interstellar extinction (central $A_V$ = 3.96 or $A_{Ks}$ = 0.47 mag,  
\citealt{Schlafly2011}) but the NIR windows present a distinctive and relatively unpolluted vantage point.  Nevertheless, the density of the foreground stellar population  presents an additional challenge.

\cite{Baravalle2018} proposed a methodology employing \textsc{SExtractor} + \textsc{PSFEx} \citep{Bertin2011} on VVV and VVVX images to identify galaxies behind the Galactic disc. The extended objects were separated from point sources that satisfied  morphological criteria based on the stellar index (\textit{CLASS\_STAR}), the SPREAD\_MODEL ($\Phi$) parameter, the radius that contains 50\% of the total flux of an object ($R_{1/2}$), and the concentration index ($C$, \citealt{Conselice2000}). The possible extragalactic sources also met photometric criteria based on $K_{s}$ magnitudes and NIR colours. This procedure has been successfully employed in the following: The VVV NIRGC I catalogue \citep{Baravalle2021}, comprising 5,554 galaxies that underwent visual inspection in the inner regions of the southern Galactic disc, and the VVV NIRGC II catalogue \citep{Daza2023}, incorporating 1,003 galaxies from the northern portion of the Galactic disc. 
In order to mitigate the occurrence of false detections, a convolutional neural network was implemented to analyse the VVVX images (IS-CNN), and a gradient boosting model was utilized on the photometric and morphological data (PS-XGBoost). The catalogue included both probability classifiers, and all galaxies were also subjected to a visual inspection. The most recent publication, the VVV NIRGC III catalogue \citep{Alonso2025}, presents a total of 167,559 galaxies that were identified in the VVVX region of the southern Galactic disc, with the probabilities associated with each galaxy being derived from machine learning algorithms. Altogether, the southern Galactic disc catalogues contain 173,113 galaxies, of which only 493 galaxies have spectroscopic redshifts, and 3,670 have photometric redshifts. Throughout this work, the galaxies from the VVV NIRGC I and III catalogues will be referred to as VVV NIRGC galaxies, and their morphological and photometric parameters will be used in the analysis.

At low Galactic latitudes,
the need for spectroscopic surveys in the NIR is becoming increasingly evident. There are some HI surveys such as MeerKAT \citep{Goedhart2024} and WALLABY \citep{Koribalski2020, Westmeier2022}, but more redshift measurements of galaxies need to be obtained in the VVV and VVVX disc regions. While spectroscopic redshifts offer the highest precision, especially in the Local Universe, their acquisition becomes extremely challenging in regions heavily obscured by the Galactic plane. 

Photometric redshift ($z_{phot}$) estimations have emerged as a powerful methodology for studying galaxies in regions where spectroscopic observations are either limited or unfeasible. 
The applicability of photometric techniques at very low redshifts is supported by previous studies. For instance, \citet{Marleau} used mid- to far-infrared surveys such as the Galactic Legacy Infrared Mid-Plane Survey Extraordinaire (GLIMPSE, \citealt{benjamin}) and
MIPS Galactic Plane Survey (MIPSGAL, \citealt{carey}), and successfully reported the discovery of 25 previously unknown galaxies at low latitudes in the Sagitta-Aquila region ($l \sim 47 -55^\circ$ and $|b|  \lesssim 1^\circ$), obtaining photometric redshifts in the $z\simeq 0.01 - 0.05$ range. This demonstrates the feasibility of this technique for mapping the galaxy distribution in the zone of avoidance. In addition, recent studies have revealed the robustness of $z_{phot}$ at low redshifts when precise calibration procedures are employed. \citet{2020A&A...636A..90S}, for example, used convolutional neural networks trained on the Sloan Digital Sky Survey (\citealt{alam}), and by incorporating the k-nearest neighbours approach from \citet{beck} and using spectroscopic redshifts from Galaxy and Mass Assembly (GAMA, \citealt{baldry}) derived $z_{phot}$ with 
median absolute deviations of $\simeq 0.009$ for $r < 17.8$ mag at a redshift of  0.1.
Similarly, \cite{cande} utilised a machine learning approach, namely the Bayesian Mixture Density Networks, with the Southern Photometric Local Universe Survey Data Release 4 (\citealt{Herpich2024}) data, and estimated photometric redshifts to identify isolated galaxy pairs with typical $z_{phot} \sim 0.04 - 0.25$, achieving uncertainties of around $\sim$ 0.01.
 
In this study, we use morphological and photometric properties of the VVV NIRGC galaxies to address the satellite population of the Circinus galaxy, employing photometric redshift estimates to discard foreground galaxies. This paper is organised as follows: Sect.~ \ref{section2} presents the VVV and VVVX data together with surveys at other wavelengths. The search for satellite galaxies using the available photometry
is discussed in Sect.~\ref{section3}. The photometric redshift procedure is described in detail in Sect.~\ref{section4}.  
The distribution of galaxies around the Circinus galaxy is addressed in Sect.~\ref{section5}, and finally, the main conclusions are highlighted in Sect.~\ref{section6}.

\section{The data}
\label{section2}

The Circinus galaxy is located at low Galactic latitudes ($b = -3.808^{\circ}$). In this region, significant obscuration due to dust and stellar density from our own Galaxy contributes with high interstellar extinction. At the Circinus distance, the virial radius of 220 kpc subtends an angular radius of approximately 3 degrees in the sky.  
This angular size is indicative of a circular region surrounding the Circinus galaxy, within which a satellite system could be found. 

\subsection{The VVV NIRGC galaxies}

\begin{figure*}
   \centering
   \includegraphics[width=1\textwidth]{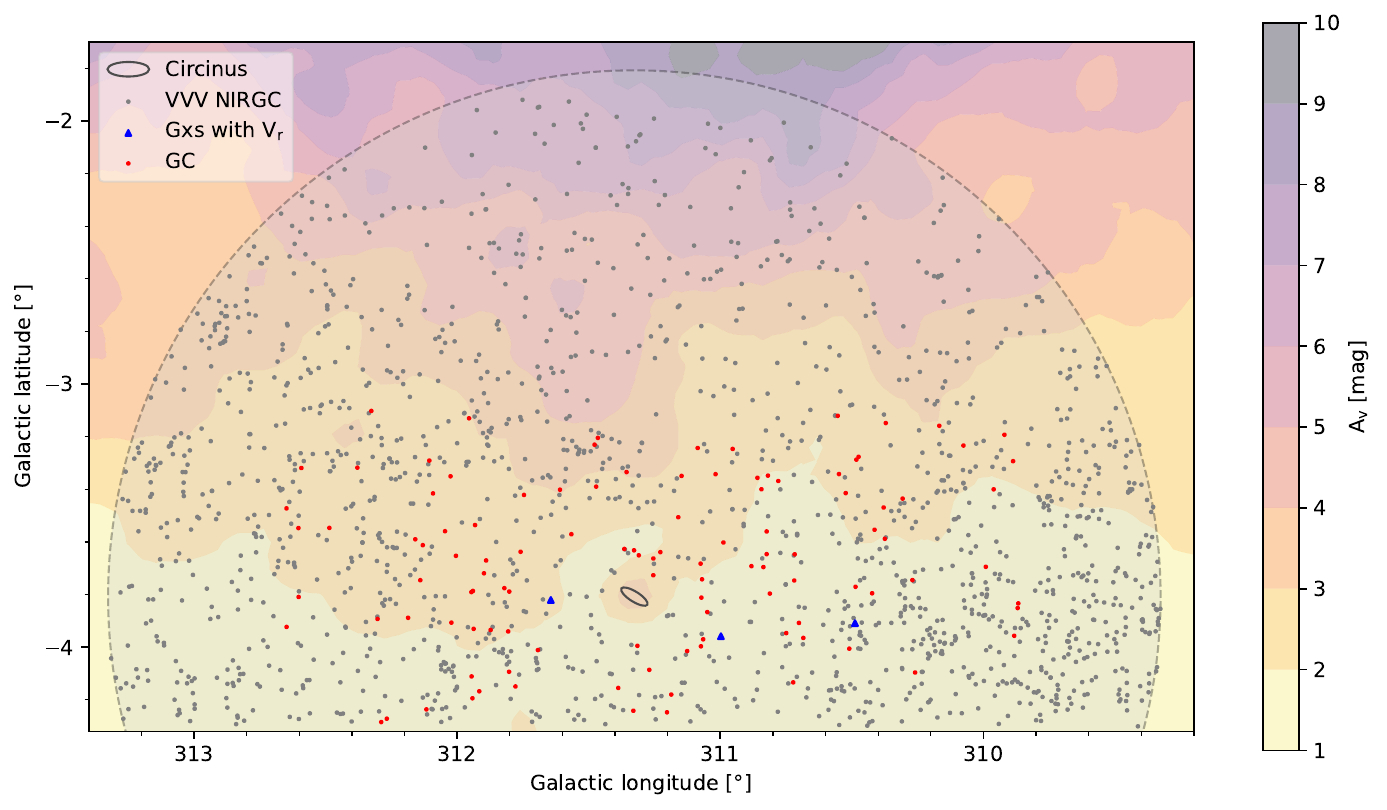}
   \caption{The distribution of the VVV NIRGC galaxies in Galactic coordinates around the Circinus galaxy. The grey dashed circle represents the studied region, and the black ellipse shows the size and orientation of the Circinus galaxy. The galaxies are shown with black circles, and the three galaxies with available spectroscopy are indicated by blue triangles. The globular cluster candidates from \cite{Obasi2023,Obasi2024} are represented with red circles.  
   }
   \label{fig:MAP1}%
    \end{figure*}

The VVV and VVVX surveys do not completely cover the circular region of a 3-degree radius around the Circinus galaxy.  Thus, as a compromise, we defined an area within a 2-degree radius of the galaxy, which is covered by a small part of the VVV survey down to Galactic latitudes of $-2.25^{\circ}$.  The remaining part is restricted to Galactic latitudes of $ b \le -4.5^{\circ}$, which corresponds to the limits of the VVVX survey.  Figure~\ref{fig:MAP1} shows this area around the Circinus galaxy in Galactic coordinates as a grey dashed circle, and throughout the present work, we will refer to this as the studied region. The background is colour-coded by the A$_V$ interstellar extinction, and the Circinus galaxy is represented by an ellipse with semi-major and minor axes of 3.5 and 1.15 arcmin, respectively, estimated from the $K_s$ image of the VVVX survey. 

The studied region represents a circular area of about 2/3 of the Circinus virial radius, which is appropriate for searching for satellites of this galaxy. We selected 26 galaxies from VVV NIRGC I and 1,516 from VVV NIRGC III.  Both catalogues include the galaxy positions  together with their NIR photometry, $JHK_s$ magnitudes and colours, and with morphological parameters such as half-light radius ($R_{1/2}$), concentration index (C), ellipticity ($\epsilon$), and the Sersic index (n, \citealt{Sersic1968}). 
In total, the 1,542  VVV NIRGC galaxies  were visually inspected using $J$, $H$, and $K_{s}$ images, as was carried out for VVV NIRGC I \citep{Baravalle2021}. The galaxies are located in reddened regions of the Galactic disc, with estimated interstellar extinction values ranging from $A_V=1.5$ to $9$ mag, equivalent to $A_{Ks}= 0.18$ to $1.06$ mag. The VVV NIRGC III \citep{Alonso2025} reaches 16.0 mag and, due to the extinction limitations, the magnitude distribution of the galaxies around Circinus reaches $K_s$ = 15.5 mag, which corresponds to a completeness of 90\% of the detections. Considering the Circinus distance of 4.2 Mpc, the limiting magnitude of 15.5 mag corresponds to an absolute magnitude $M_{K_s} = -12.6$ mag. Therefore, we can study the satellite population expected to be found around the Circinus galaxy.

A search was conducted for previously known galaxies in the studied region around the Circinus galaxy using the SIMBAD astronomical database \citep{Wenger2000}. A total of 21 galaxies were identified, with the exclusion of a few duplicate entries and some background QSOs.
Only seven of these galaxies have estimated radial velocities, of which the following three are included in the VVV NIRGC III catalogue: LEDA 166334 (2999 $\pm$ 70 km/s), ESO 97-12 (3127 $\pm$ 92 km/s), and 2MASX J14054516-6541032 (3024 $\pm$ 157 km/s) from \cite{Fairall1998}. These galaxies are located at a significant distance from the Circinus galaxy, whose systemic velocity is 439 km/s \citep{Jones1999}. We also searched for galaxies with available radial velocities in the complete area of the 2-degree radius around the Circinus galaxy that is not covered by the VVV and VVVX surveys. There are 4 additional galaxies reported by \cite{Staveley2016} within the radial velocity range of 1647 to 8490 km/s clearly distant from the Circinus galaxy. 

Figure~\ref{fig:MAP1} also shows the distribution of the selected 1,542 VVV NIRGC galaxies 
in the studied region around the Circinus galaxy. 
These galaxies are represented by black circles, with the only three galaxies with available radial velocities indicated by blue triangles. The 111 globular cluster candidates studied by \cite{Obasi2023,Obasi2024} are shown as red circles. 
A discernible interstellar extinction gradient as a function of Galactic latitude is evident, with a general decrease in the number of galaxies towards the inner parts of the Galactic disc, which is also a consequence of increasing interstellar extinction.
On examining the A$_V$ interstellar extinctions versus the NIR magnitudes of VVV NIRGC galaxies, it is clear that the sample of galaxies becomes progressively more incomplete with increasing interstellar extinction. The spatial distribution of galaxies is non-uniform, with clear empty regions and some clustered objects. Despite the identification of a few galaxy groups at these low latitudes, the distribution does not solely reflect genuine clustering but also the impact of differential reddening.

\subsection{Galaxies with photometry in optical, near- and mid-IR wavelengths}

The Dark Energy Camera Galactic Plane Survey (DECaPS2, \citealt{Schlafly2018}) is a five-band optical and NIR survey of the southern Galactic plane with the Dark Energy Camera on the 4.0 m Blanco telescope, covering the very faint magnitudes: $g<23.5, r<22.6, i<22.1, z<21.6$, and $Y<20.8$ mag \citep{Saydjari2023}. We cross-matched our galaxies in the region of the Circinus galaxy, and identified 1,541 galaxies in common using an angular separation of 1 arcsec. Only 1,039 (67\%) of these galaxies have photometry in the five passbands, and more than 90\% have some NIR magnitudes. Of the total, 1,085 galaxies (70\%) have $g$ magnitudes located in regions with lower interstellar extinctions.

The Gaia satellite of the European Space Agency \citep{Gaia2016} has released an updated large optical survey, Gaia Data Release 3 (Gaia DR3, \citealt{Brown2021}), which provides homogeneous astrometry and photometry for sources with $G < 21$ mag across the entire sky. In particular, Gaia DR3 contains some automatic classifiers, such  as galaxies and QSOs in the database \citep{Delchambre2023}.
It is not expected that Gaia detects extended sources as galaxies, but more compact and concentrated galaxies that are sufficiently bright should match well with our galaxies. Exploring the studied region around the Circinus galaxy, we identified 323 VVV NIRGC  galaxies that cross-match with the Gaia DR3 database using an angular separation of 1 arcsec. 
There is a good overlap of the matched sources that are brighter and bluer, while Gaia misses a large number of the fainter and redder sources, as expected.

The Wide-field Infrared Survey Explorer mission (WISE; \citealt{Wright2010}) mapped the whole sky in the mid-IR: 3.4, 4.6, 12, and 22 $\mu$m passbands.  
The VVV NIRGC galaxies were cross-matched with WISE data with an angular separation of 1 arcsec.  In the area surrounding the Circinus galaxy, there are 177 galaxies with WISE counterparts. For these galaxies, the median half-light radius is R$_{1/2}$ = 1.32 $\pm$ 0.90 arcsec, a larger value than that found for the rest of the galaxies in the studied region, which is R$_{1/2}$ = 0.83 $\pm$ 0.39 arcsec. Also, these galaxies are one magnitude brighter for $K_{s}$  magnitudes, with a median of 13.94 $\pm$ 1.10 mag.


\section{Searching for Circinus satellites with the available photometry}
\label{section3}

\subsection{Colour-colour diagrams}

\begin{figure*}
   \centering
   \includegraphics[width=0.49\textwidth]{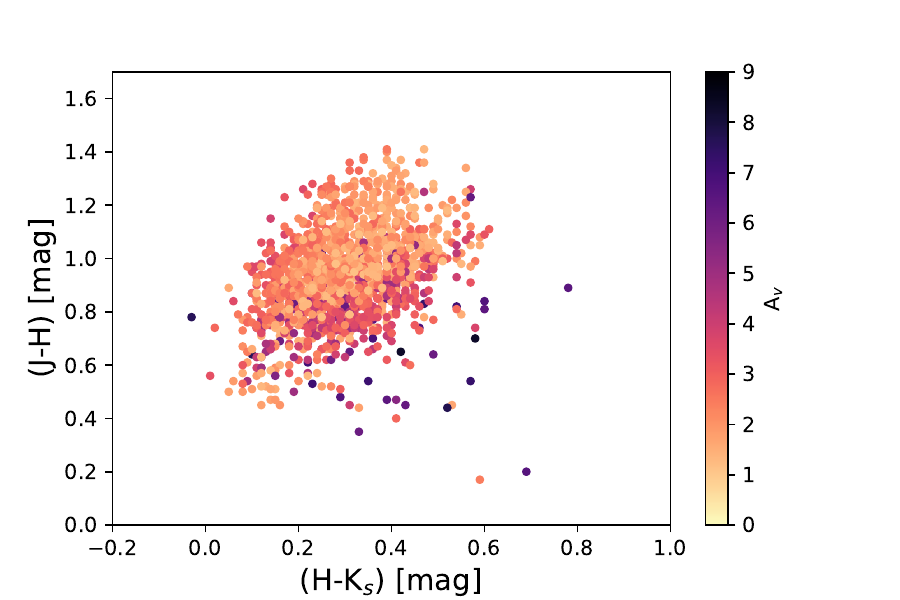}     \includegraphics[width=0.49\textwidth]{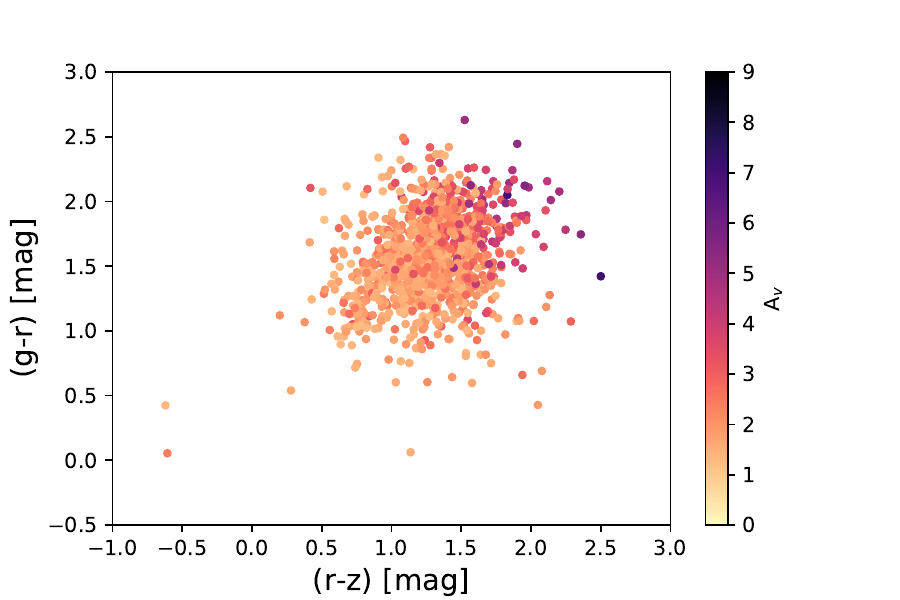}
   \caption{Colour-colour diagrams of the VVV NIRGC galaxies around the Circinus galaxy 
   using the VVV and VVVX photometry (left panel) and DECaPS2 survey (right panel). The auxiliary bar represents the A$_V$ interstellar extinctions.}
              \label{fig:CMD2}
    \end{figure*}

Figure~\ref{fig:CMD2} shows the NIR colour-colour diagram (CCD) for the 1,542 VVV NIRGC galaxies selected in the studied region around the Circinus galaxy (left panel). The optical CCD for the 1,085 VVV NIRGC galaxies with $g$ magnitudes from DECaPS2 is displayed in the right panel. In general, there are no clear tendencies based on these diagrams. However, higher dispersions are found for galaxies 
located in regions with higher interstellar extinctions, with $g$ magnitudes being more affected. 

Mid-IR WISE colours are often used to select galaxies with active nuclei \citep{Jarrett2011,Stern2012,Assef2018}. These galaxies present a redder (3.4 - 4.6) colour for non-active galaxies since warm dust emission is more determinant than the light coming from the old stellar population in the host galaxy \citep{Ceccarelli2022}. The (4.6 - 12) colour can also be used to identify star-forming and passive galaxies, with values of (4.6 - 12) $>$ 3 mag and (4.6  -12) $<$ 1.5 mag, respectively \citep{Jarrett2017}. Using these criteria, of the 177 NIRGC galaxies with WISE data,
we identified 85 star-forming (SF) galaxies and 33 passive (P) ones around the Circinus galaxy.
Moreover, six galaxies satisfy the criteria of \cite{Assef2018} and \cite{Stern2012} to be QSO candidates, while one galaxy falls within the box defined by \cite{Jarrett2011} as an AGN candidate.

\subsection{Sizes and structural parameters}

Within the Local volume, satellite galaxies are generally divided into dwarf satellites with radii of 0.05 to 0.5 kpc and larger satellites with radii of 0.5 to 1.5 kpc \citep{Muller2020}. 
At the distance of the Circinus galaxy, 
these physical scales translate to an angular radius of 2.45 to 24.58 arcsec for dwarf satellites and 24.68 to 73.74 arcsec for larger satellites, thereby providing an observational reference for potential satellite candidates. 

\begin{figure}
   \centering
\includegraphics[width=0.52\textwidth]{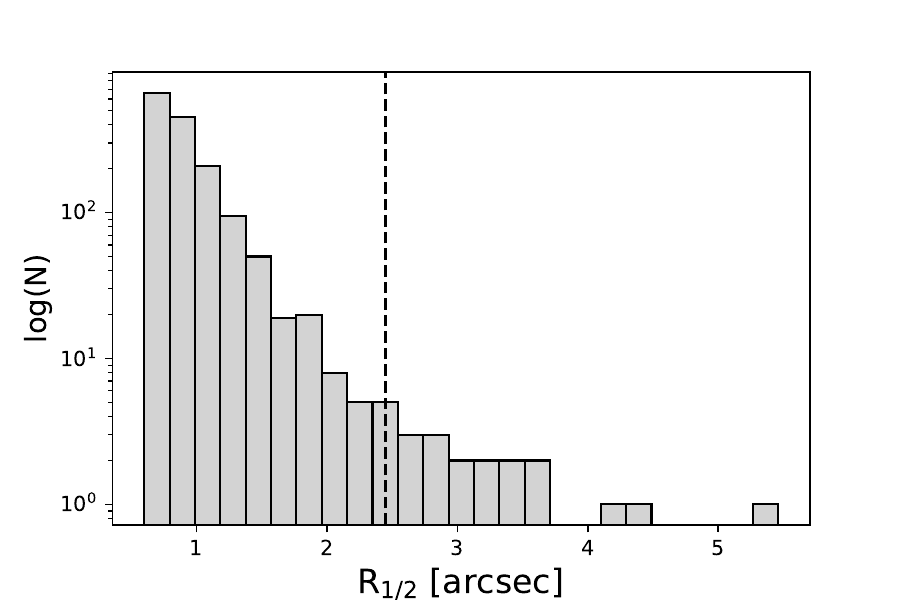}
   \caption{Half-light radius distribution of the galaxies around the Circinus galaxy in logarithmic scale. The black dashed line shows the lower limit in half-light radius for dwarf satellites at the distance of the Circinus galaxy. 
   }
     \label{fig:Rdistribution}
    \end{figure}

Figure~\ref{fig:Rdistribution} shows the half-light radius distribution for VVV NIRGC galaxies around the Circinus galaxy. 
The lower limit of typical dwarf satellites in the Local volume in arcseconds, at the Circinus distance, is included in the figure. %
Although the majority of objects are likely to be background galaxies, 20 galaxies exhibit a half-light radius in the range of 2.45 $< R_{1/2} <$ 5.46 arcsec. These angular sizes are comparable to the typical physical sizes of dwarf galaxies with radii of 0.05 to 0.5 kpc, which could plausibly represent satellite candidates of the Circinus galaxy. Of these candidates, two were classified as star-forming galaxies (VVVX-J135433.49-653910.2 and VVVX-J141829.93-643221.4) and one as a passive galaxy (VVVX-J140339.44-653516.1). 

There is a wide range of ellipticities measured for galaxies of all magnitudes and sizes, although the vast majority of the most concentrated galaxies tend to be rounder. The VVV NIRGC galaxies also possess a wide range of Sersic indices, which are comparatively independent of their NIR magnitudes. Table~\ref{tab:statistics} shows the median values of the extinction-corrected $K_s$ magnitudes and colours and the morphological parameters R$_{1/2}$, $C$, $\epsilon$, and $n$ for the 1,542 galaxies around the Circinus galaxy and for the 85 SF and 33 P galaxies defined with the WISE data.
We did not find any clear trend between the morphological parameters of the galaxies and the angular distance to the Circinus galaxy.

\begin{table}
\small
\caption{Median photometric and structural parameters of the galaxies around Circinus.}
\begin{tabular}{lccc}
\hline
\hline
Parameter &  All        & SF       & P \\
          &   galaxies  & galaxies & galaxies \\
\hline
$K^0_s$ (mag)       &  15.01 $\pm$ 0.39 & 14.19 $\pm$ 0.92 &  13.14 $\pm$ 0.97 \\
$(J - K_s)^0$ (mag) &   1.23 $\pm$ 0.23 & 1.20 $\pm$ 0.23 &   1.11 $\pm$ 0.17\\
$(H - K_s)^0$ (mag) &   0.30 $\pm$ 0.11 & 0.25 $\pm$ 0.10 &   0.24 $\pm$ 0.83\\
$R_{1/2}$ (arcsec)  &   0.83 $\pm$ 0.39 & 1.17 $\pm$ 0.65 &   1.21 $\pm$ 0.51\\
C                   &   2.48 $\pm$ 0.29 & 2.58 $\pm$ 0.33 &   2.86 $\pm$ 0.38 \\
$\epsilon$          &   0.24 $\pm$ 0.15 & 0.32 $\pm$ 0.18 &   0.21 $\pm$ 0.11 \\
n                   &   4.51 $\pm$ 0.97 & 3.60 $\pm$ 1.83 &   5.40 $\pm$ 1.78\\
\hline
\end{tabular}
\label{tab:statistics}
\end{table}

\section{Photometric redshifts using VVV and VVVX data}
\label{section4}

Photometric redshift estimations have become an essential tool for investigating galaxies in areas of the sky where 
spectroscopic data are scarce. 
This is particularly relevant at low Galactic latitudes, 
where the VVV and VVVX surveys facilitated the detection of a significant number of extragalactic objects and provided their NIR photometry.
Estimating redshifts for the galaxies in the studied region around the
Circinus galaxy allows us to explore their distribution. 
In these regions, although spectroscopic redshifts are unavailable, well-calibrated photometric methods
provide a valid initial step for filtering and prioritising candidate
galaxies. Our specific method employs traditional machine
learning with $JHK_s$ photometry and local training.

In order to estimate redshifts, it is necessary to have a sample of galaxies that possesses consistent and complete photometric coverage, as well as spectroscopic information.  
Cross-matching with other surveys, such as DECaPS2, Gaia, and WISE, yielded valuable multi-wavelength data. However, it was found that only a small fraction (less than $1\%$) of the galaxies had simultaneous detections, for example, in the three NIR passbands of the VVV and VVVX surveys and all five DECaPS2 optical passbands.
This limitation led to a substantial reduction in the statistical representativeness of the expanded multi-passband dataset.  Consequently, we chose to prioritise the analysis of the internal photometric data from the VVV and VVVX surveys. By focusing on the $J$, $H$, and $K_{s}$ passbands, we were able to retain a substantially larger sample with homogeneous coverage, which is crucial for training a robust and reliable machine learning model. This strategy allowed us to estimate photometric redshifts for the 1,542 galaxies in the studied region, preserving the statistical power of the sample while ensuring consistency in the input features.

\subsection{Methodology and the use of the ANNz Code}

The Artificial Neural Network for Photometric 
Redshifts (ANNz V2.3.1, \citealt{2016PASP..128j4502S}) code is a machine learning tool 
specifically designed for photometric redshift estimations. 
The neural network model is trained using a sample of galaxies with known spectroscopic redshifts 
and photometric measurements. Once trained, this 
model can predict the redshift of galaxies based solely on their photometric properties.

In order to train the ANNz model for the VVV and VVVX data, a sample of 486 galaxies with known spectroscopic redshifts, and photometry
in the $J$, $H$, and $K_s$ passbands was used. 
These passbands are crucial for capturing the 
relevant spectral information needed for
accurate redshift estimations. 
To ensure a reliable model, the sample was divided into three subsets: training (228 galaxies), validation (142 galaxies), and testing (116 galaxies). 
This division helps prevent overfitting and allows for a robust evaluation of the model's performance.  The calibration of the subsets is used to train the neural network, adjusting its parameters to minimise the errors in redshift predictions.  The validation set is used to monitor the model's performance during training, helping to avoid overfitting and tuning hyper-parameters. The test set is then used to evaluate the final model's predictive ability, thereby providing an estimate of the error in the predicted photometric redshifts.

The combination of the three $J$, $H$, and $K_{s}$ passbands,
along with their corresponding colour combinations ($J-K_s$, $H-K_s$, and $J-H$), 
has been shown to provide
additional spectral information, improving the 
model's ability to distinguish between galaxies
with different properties, especially distant ones.
Consequently, the model can accurately estimate the 
photometric redshifts. 
The resulting ANNz architecture adopted here, is 6:12:12:12:1,  which used an input layer of six parameters: three NIR magnitudes and three colours. This feeds into three hidden layers, each of which processes data, using two different mathematical methods to contribute 12 units per method per layer (12, 12, 12). A single output layer then provides the final result.

\begin{figure}
   \centering
 \includegraphics[width=0.49\textwidth]{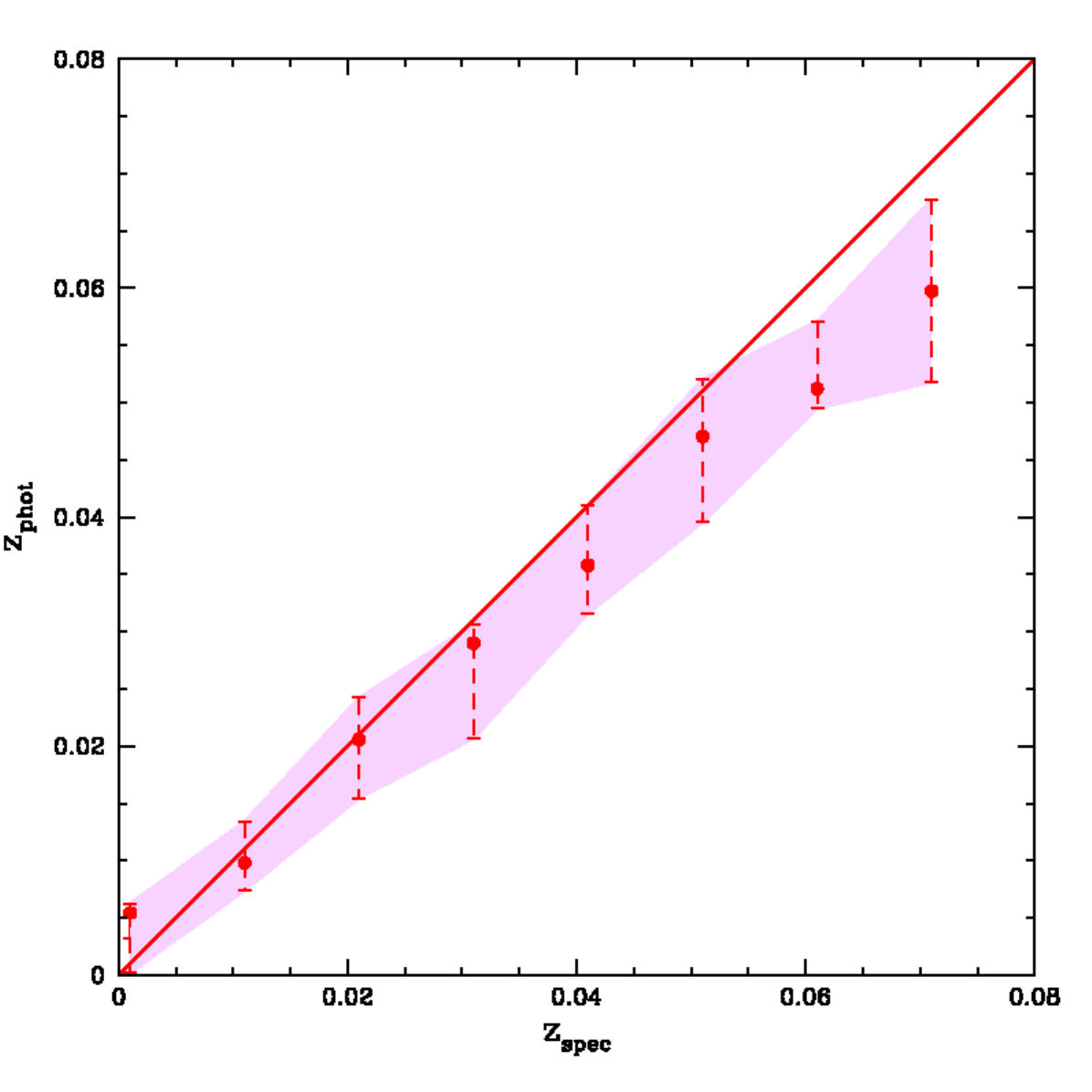}
   \caption{Calibration results for the ANNz code. 
   The solid line indicates the one-to-one relation between spectroscopic and photometric redshifts. The red points represent the mean photometric redshift within each spectroscopic redshift bin, with error bars indicating the standard deviation. The pink shaded region shows the calculated uncertainties in the distribution.}
     \label{fig:zp}
    \end{figure}

Figure~\ref{fig:zp} shows the calibration results of the ANNz code. The solid line represents the one-to-one relation between spectroscopic and photometric redshifts. The highest degree of agreement is observed in the redshift range $0.001 < z < 0.023$, where the scatter and uncertainties are significantly reduced. This improvement in performance can be attributed primarily to the larger number of galaxies with spectroscopic redshifts available in this redshift range, thereby enhancing the training and reliability of the photometric redshift estimates.  
However, for $z < 0.001$, the number of spectroscopic sources drops significantly.
Our study shows that this technique is valuable
for estimating photometric redshifts in regions of high interstellar extinction, thereby opening up the possibility of studying
galaxies without spectroscopic measurements. 

%

\subsection{Photometric redshift results}

\begin{figure}
   \centering
 \includegraphics[width=0.52\textwidth]{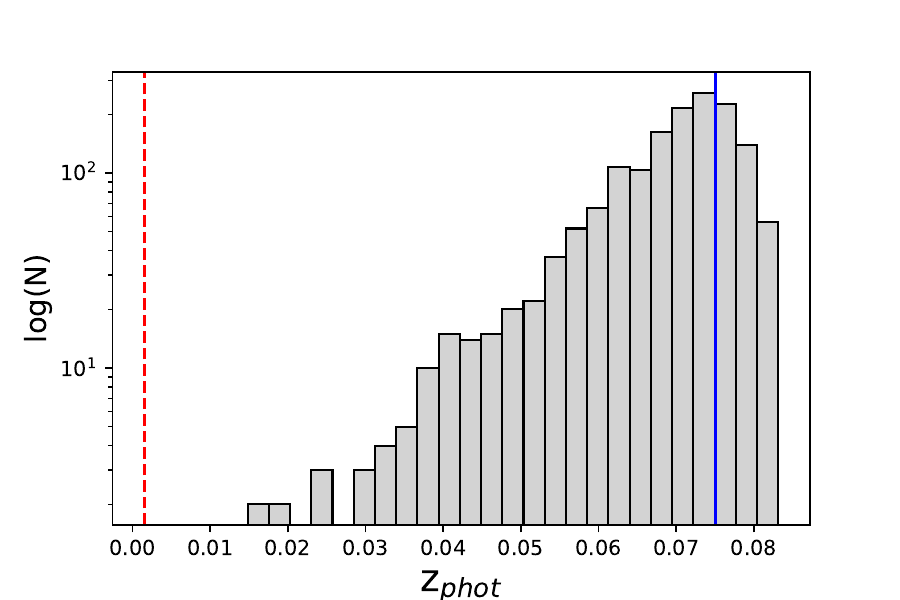}
   \caption{Photometric redshift distribution of the galaxies around the Circinus galaxy in logarithmic scale. The blue line represents the median z$_{phot}$ of the distribution and the dashed red line shows the Circinus redshift. 
   }
     \label{fig:zpdistribution}
    \end{figure}

We characterised the photometric redshift distribution of galaxies around the Circinus galaxy and quantify contamination from foreground and background sources. Figure~\ref{fig:zpdistribution} shows the distribution of photometric redshifts obtained for the 1,542 VVV NIRGC galaxies in the studied region around the Circinus galaxy. The median value of the distribution is $0.075 \pm 0.012$, and most of the galaxies are clearly background galaxies, with redshifts that extend well beyond that of the Circinus galaxy.
Only four galaxies in the sample have estimated photometric redshifts lower than $z_{\mathrm{phot}} = 0.02$, which are the galaxies with spectroscopic data. 

\begin{figure}
   \centering
 \includegraphics[width=0.56\textwidth]{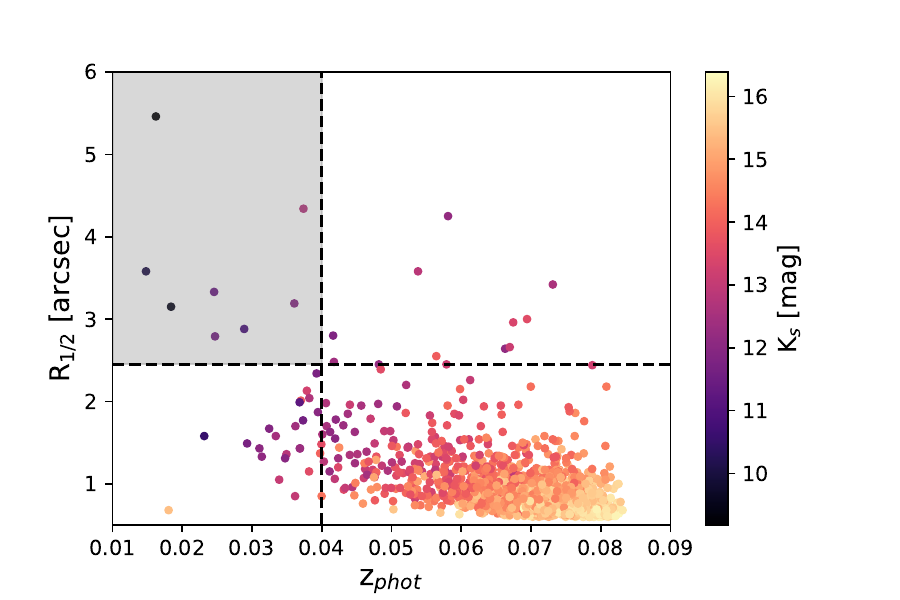}
   \caption{Half-light radii and photometric redshifts for the galaxies around the Circinus galaxy. The dashed black lines show the limit in R$_{1/2}$ for candidate satellites and an upper limit in photometric redshifts. The grey region delimited by the two lines defines possible Circinus dwarf satellites.}
     \label{fig:Rvsz}
    \end{figure}
    
\begin{figure*}
    \centering
   \includegraphics[width=0.24\textwidth]{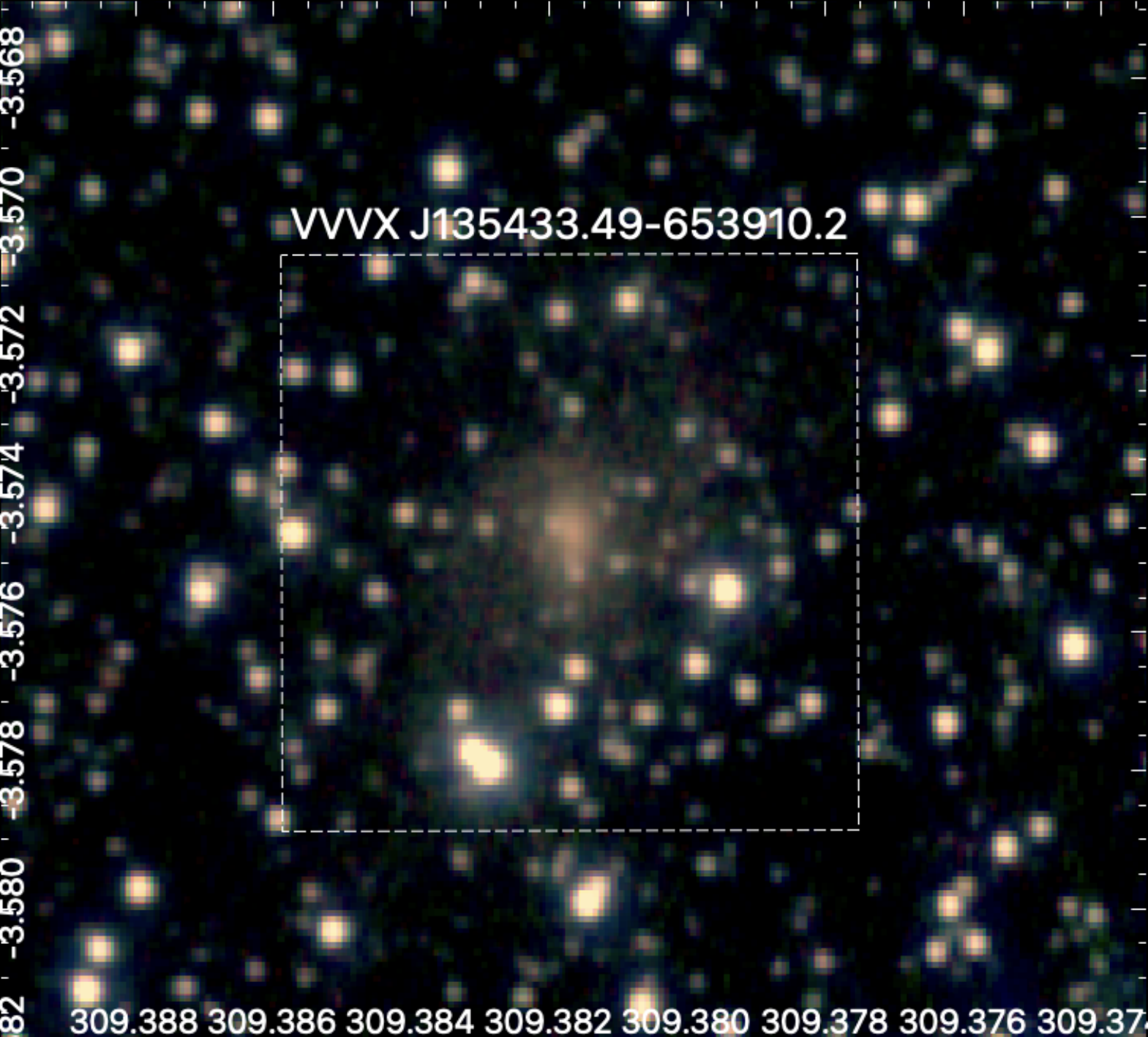}
   \includegraphics[width=0.24\textwidth]{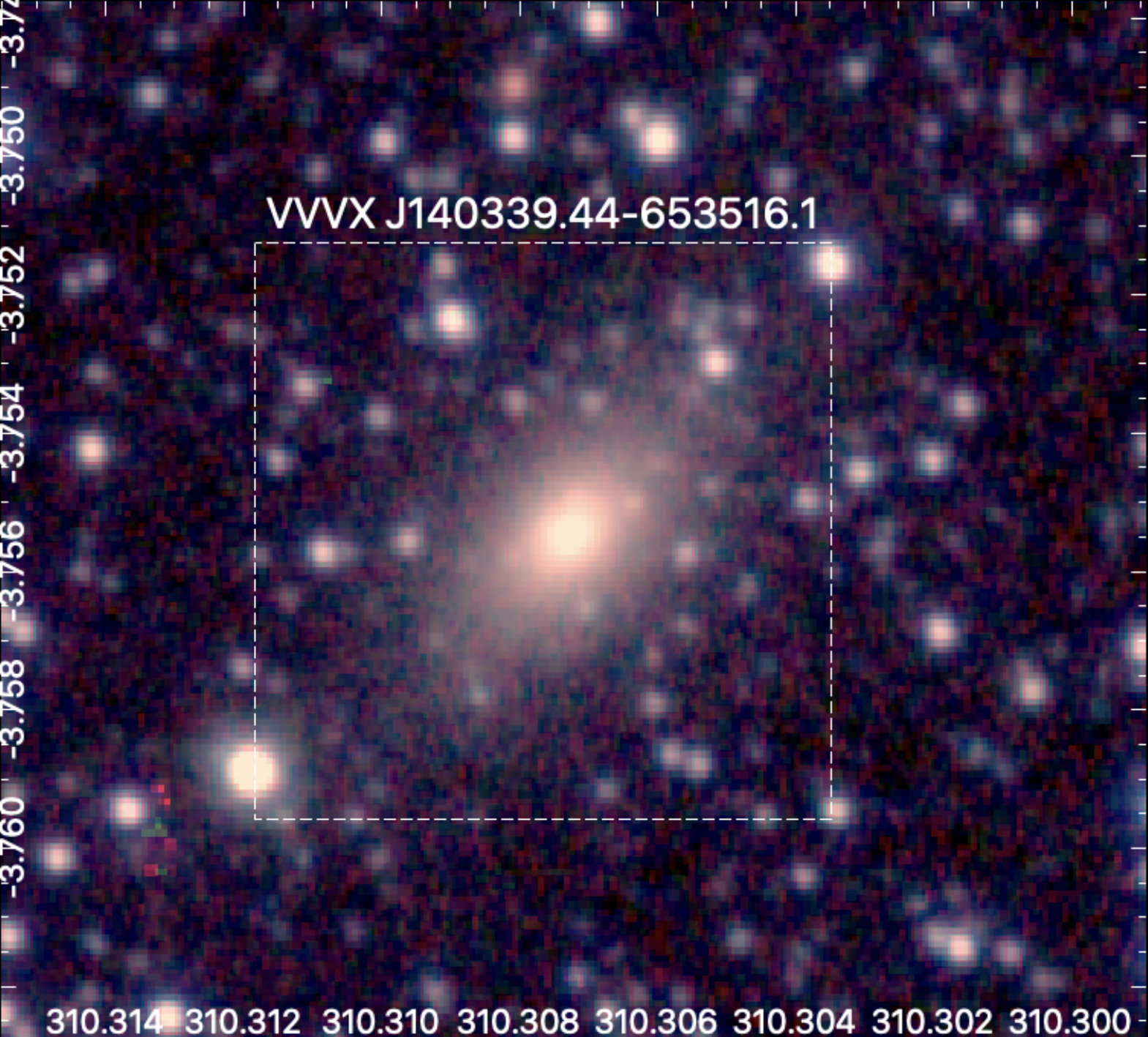}
   \includegraphics[width=0.24\textwidth]{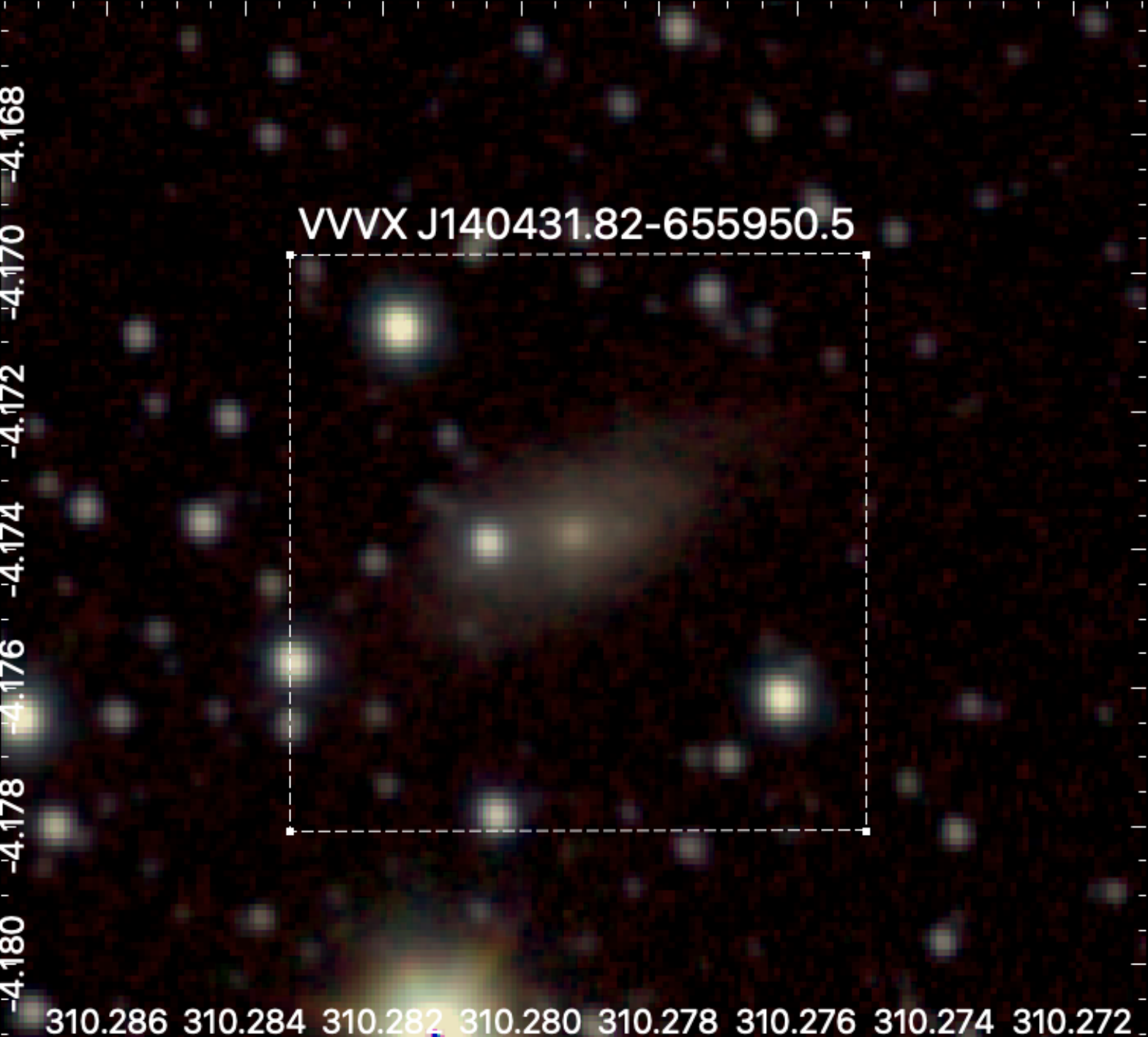}
   \includegraphics[width=0.24\textwidth]{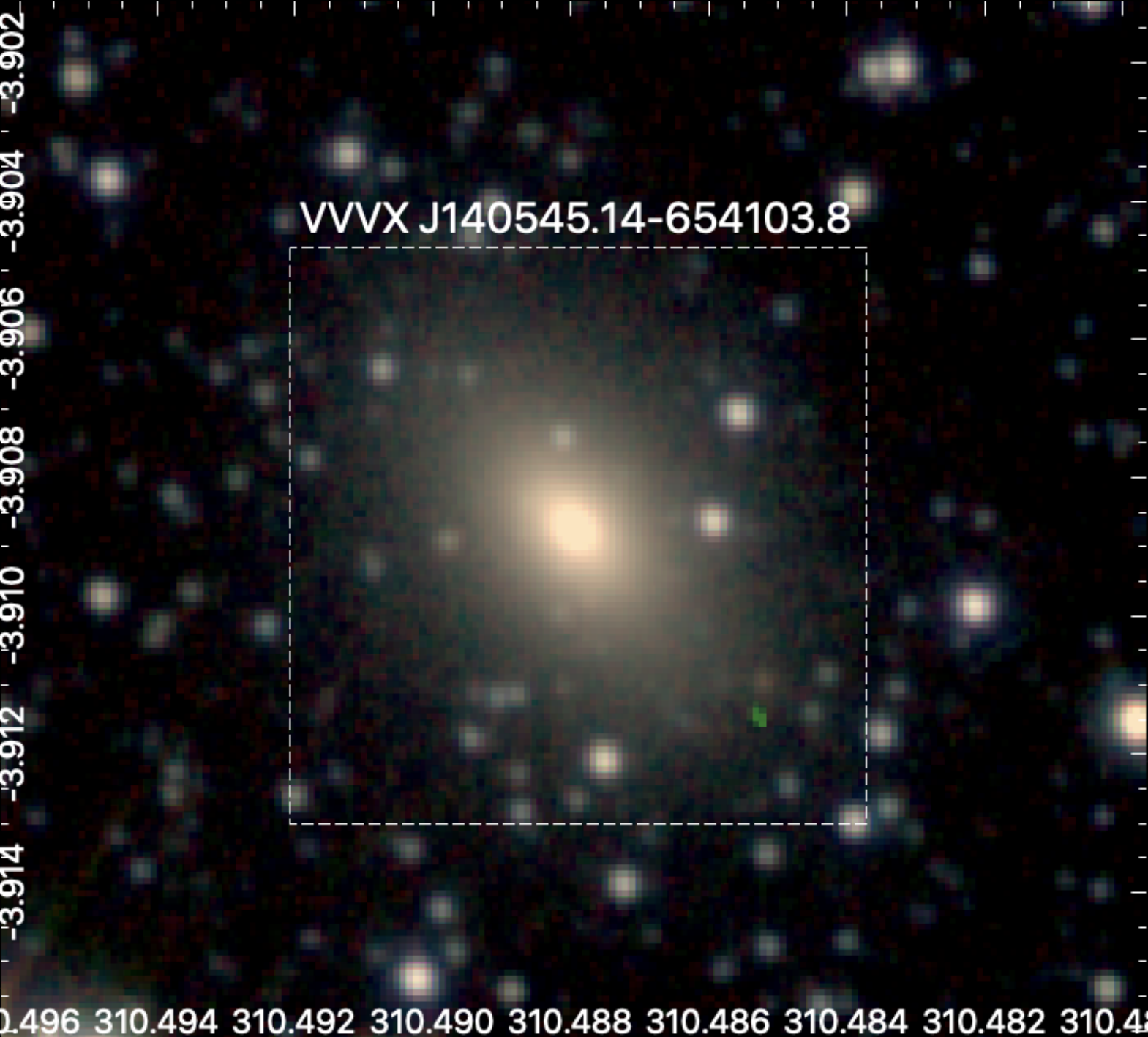}\\
   \includegraphics[width=0.24\textwidth]{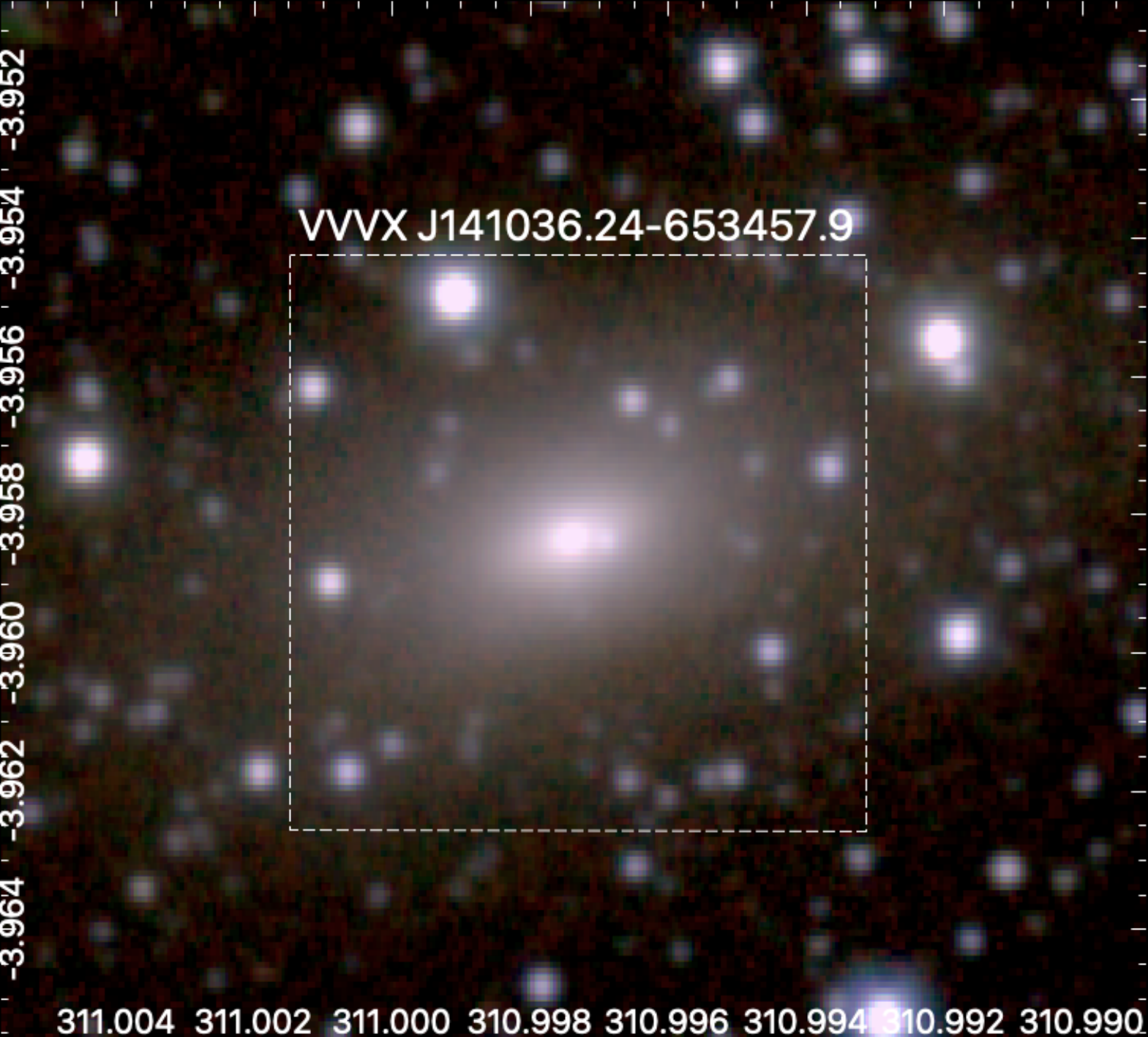}
    \includegraphics[width=0.24\textwidth]{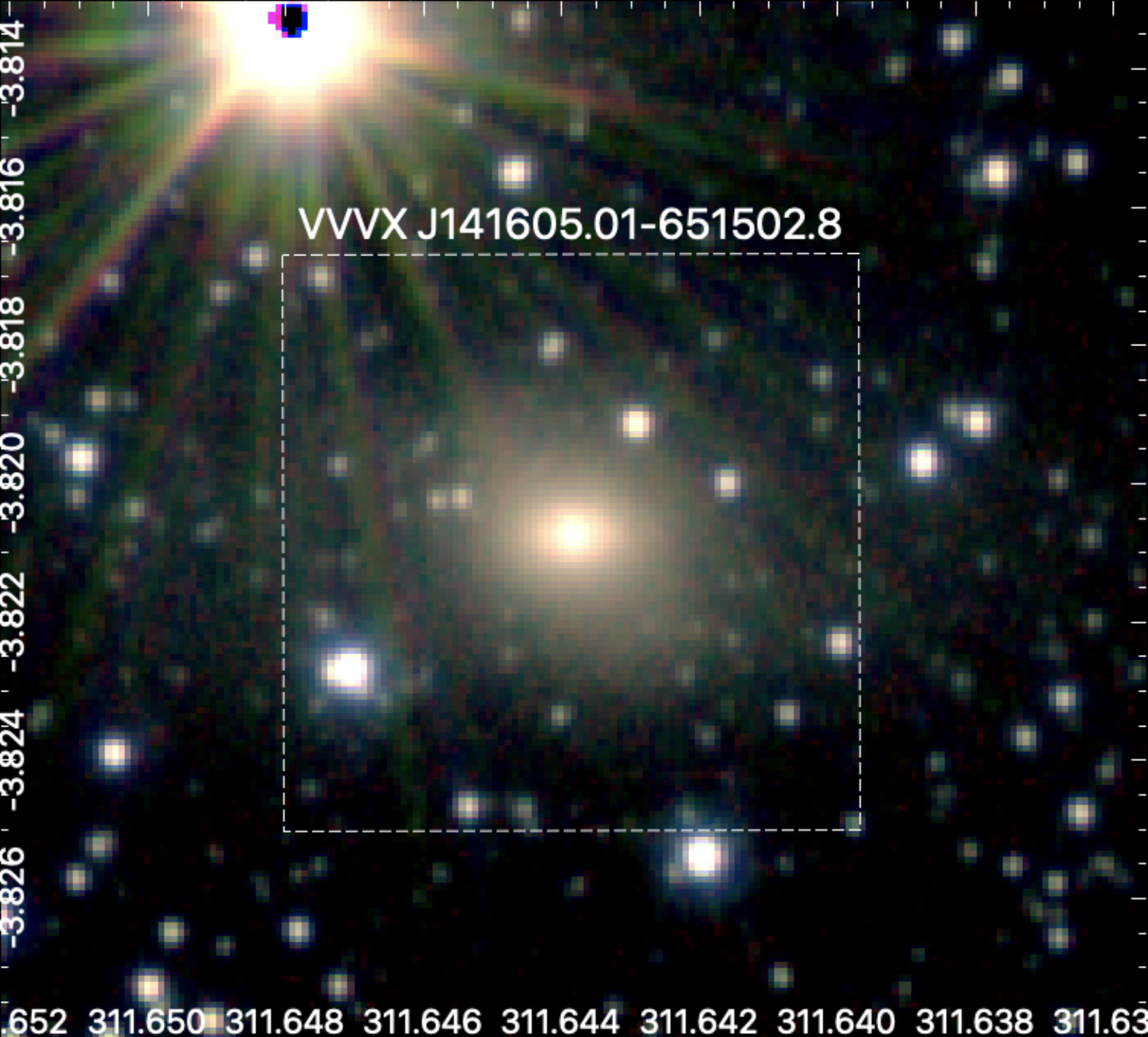}
   \includegraphics[width=0.24\textwidth]{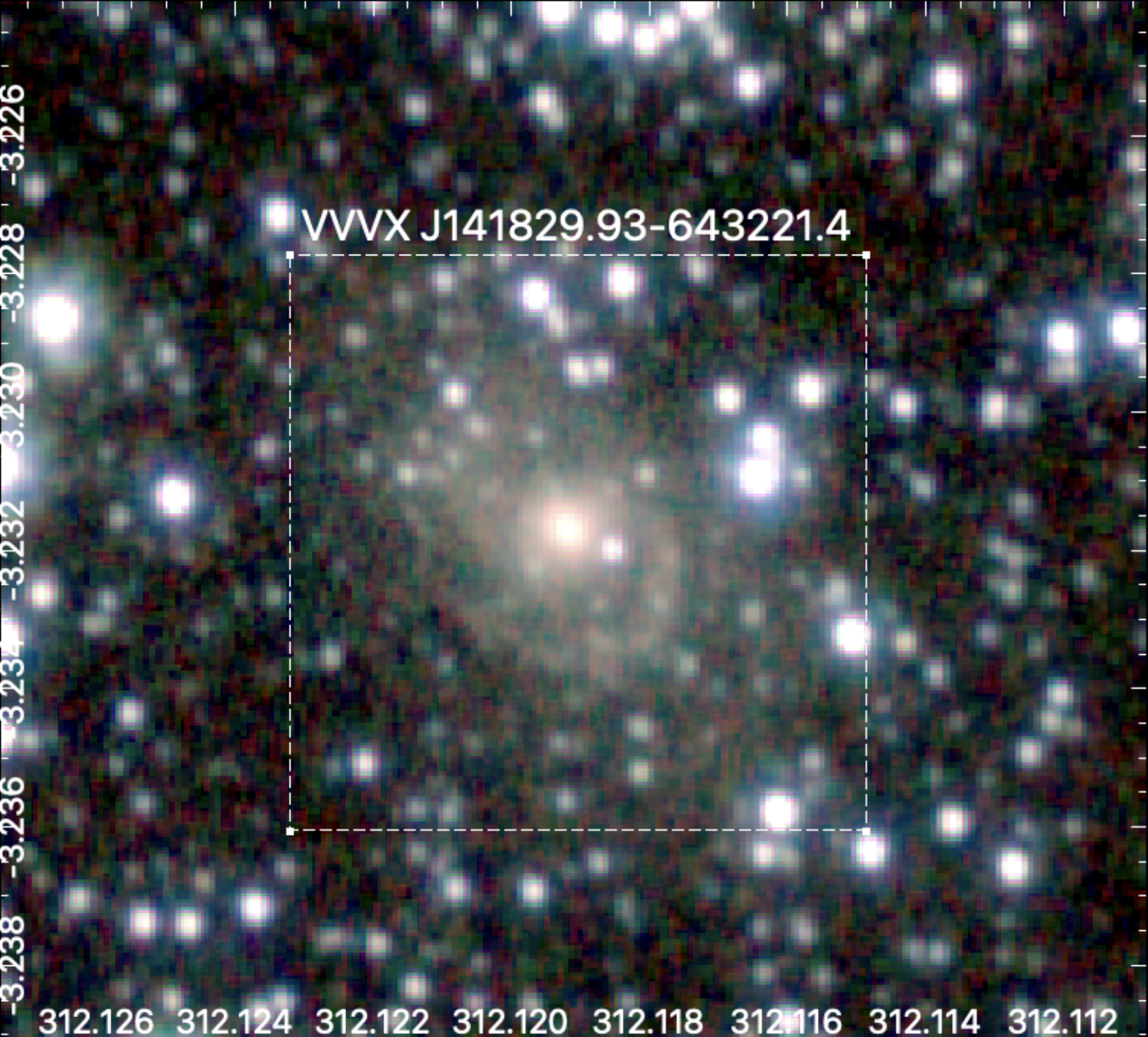}
   \includegraphics[width=0.24\textwidth]{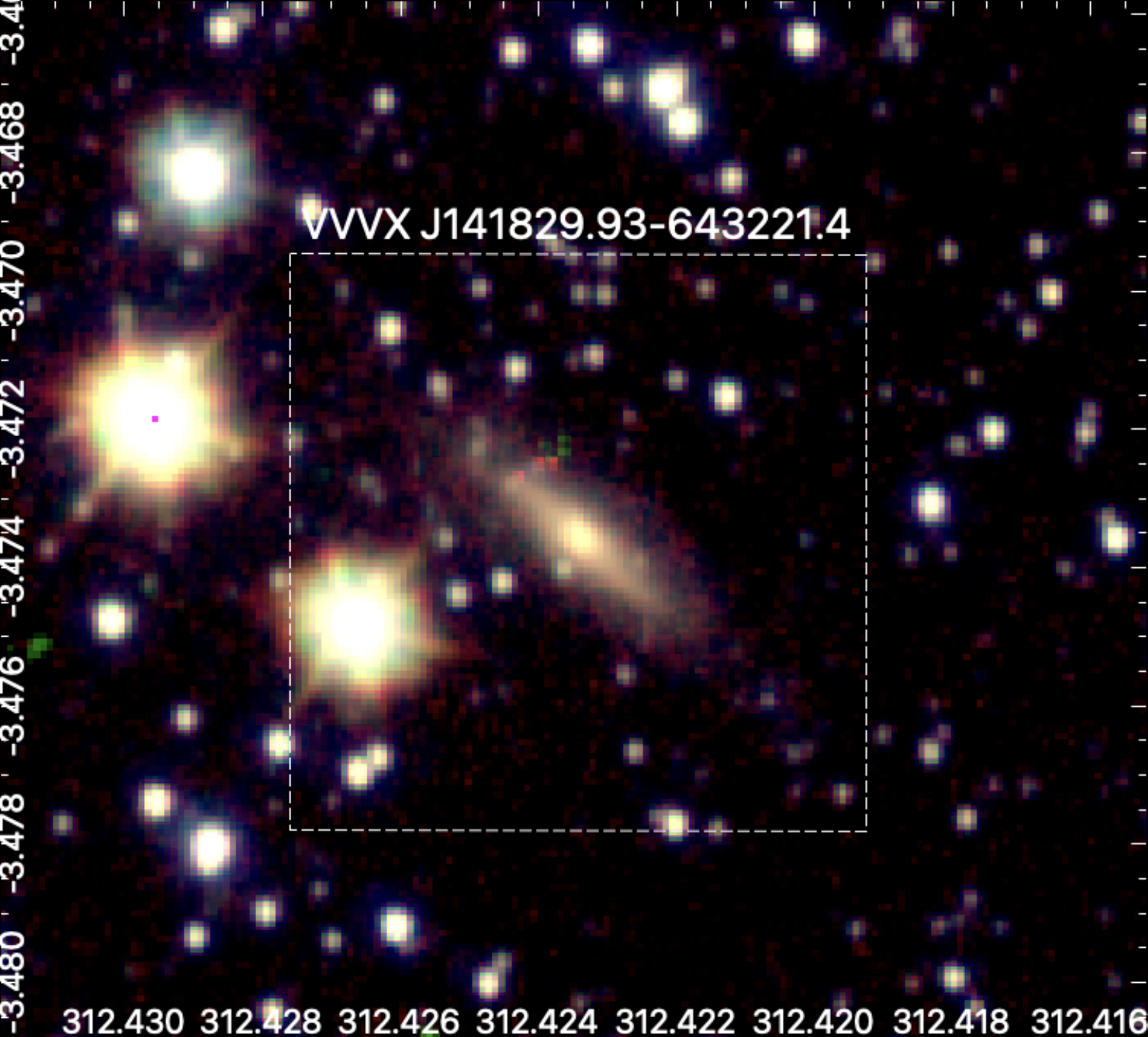}
    \caption{$K_s$ (red), $H$ (green), and $J$ (blue) colour composite postage stamps of the potential satellites of the Circinus galaxy selected from angular size and photometric redshift. These are ordered from left to right and top to bottom, following the listing order in Table~\ref{tab:satellites}. The white dashed-line box represents the field-of-view of 30 $\times$ 30 arcsec. North is up, and east is to the left.} 
    \label{fig:stamps}
    \end{figure*}

Figure~\ref{fig:Rvsz} shows the half-light radii versus photometric redshifts for the galaxies around the Circinus galaxy, colour-coded by the $K_s$ magnitudes. The black dashed lines correspond to the R$_{1/2}$ limit for candidate satellites at the distance of the Circinus galaxy, and about $\sim 3\sigma$ below the median photometric redshift. The grey region defined by the lines is the one with possible satellite candidates, which are brighter and larger VVV NIRGC galaxies. 
Of the 20 galaxies with a half-light radius in the range of 2.45 $< R_{1/2} <$ 5.46 arcsec, there are only eight with $z_{\mathrm{phot}} < 0.04$. 
Table~\ref{tab:satellites} shows the properties of the eight possible Circinus satellite galaxies selected from the angular size and smaller photometric redshift. These properties include magnitudes,  morphological parameters, and the estimated photometric redshift. The uncertainties in z$_{phot}$ are $\sim \pm 0.005$ in all cases. Figure~\ref{fig:stamps} shows the colour-composed images of these galaxies where it can be seen that they do not look like typical dwarf satellites.

\begin{figure}
   \centering
\includegraphics[width=0.52\textwidth]{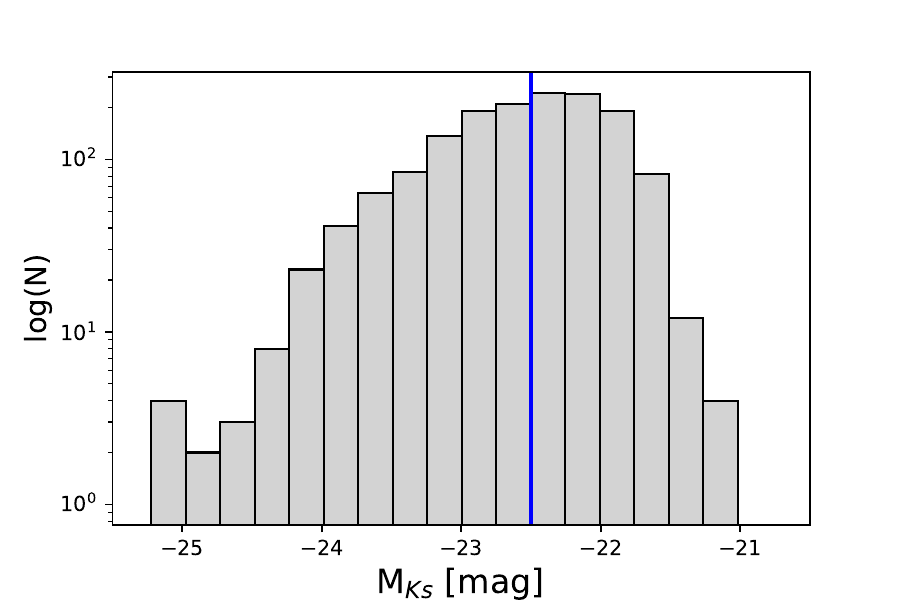}
   \caption{Absolute $K_s$ magnitude distribution of the galaxies around the Circinus galaxy in logarithmic scale. The blue line represents the median absolute $K_s$ magnitude of the distribution.}
     \label{fig:histoabsolutas}
    \end{figure}

Figure~\ref{fig:histoabsolutas} shows the absolute magnitude distribution in the $K_s$ passband ($M_{K_s}$) for the 1,542 VVV NIRGC galaxies derived using the estimated photometric redshifts of the galaxies around the Circinus galaxy. The median value is -22.50 $\pm$ 0.64 mag, represented by the blue line. Table~\ref{tab:luminosityrhalf} shows the median, minimum, and maximum absolute
$K_s$ magnitudes, together with half-light radii for the possible satellites of the Circinus galaxy: for the 20 galaxies with R$_{1/2}$ $>$ 2.45 arcsec, equivalent to 0.05 kpc, as well as the 8 galaxies with $z_{phot}$ $<$ 0.04. The latter are in the range of -25.05 to -23.39 in absolute $K_s$ magnitudes and 2.79 to 5.46 arcsec in half-light radius, which are translated into physical units of 0.05 to 0.11 kpc. These objects are far from the regions where the dwarf satellites are located in the size--luminosity phase space \citep{Mutlu2021,Doliva2025}. Furthermore, based on the photometric redshifts estimated in this work, no candidates were found that could be considered satellite galaxies of the Circinus galaxy.

This study demonstrates the potential of this approach to probe
galaxy populations in regions where optical 
spectroscopy is severely limited by interstellar extinction and 
stellar crowding. Although the uncertainties in the photometric redshifts are high, with values of about 0.005, the galaxy with
the lowest photometric redshift presents a $z_{\mathrm{phot}} =$  0.0148, which
is far from the Circinus galaxy (at z = 0.0015). 

\begin{sidewaystable*}
\small
\caption{Properties of the potential satellites of the Circinus galaxy}
\begin{tabular}{lcccccccccccccccc}
\hline
\hline
 VVV NIRGC ID & RA &  Dec & $l$ & $b$ & A$_{Ks}$ & $J^0$  &  $H^0$ & $K_s^0$ &  $J_2^0$ &    $H_2^0$ &  $K^0_s{}_2$ & $R_{1/2}$ & C &  $e$  & $n$   & $z_{phot}$  \\
 & (J2000) &  (J2000) & [$^{\circ}]$ & [$^{\circ}$] &  [mag]  &  [mag] &  [mag]  &  [mag]  &   [mag] &   [mag] &   [mag]    &  [arcsec]  &   &       &     &      \\
\hline
\hline
 VVVX-J135433.49-653910.2 & 13:54:33.49 & -65:39:10.2 & 309.382 & -3.575 & 0.22 & 13.08 & 12.79 &  12.69 &   16.52  &  15.63 &   15.43 &  4.34 & 2.71 & 0.01 &  2.87 &  0.0374   \\
 VVVX-J140339.44-653516.1 & 14:03:39.44 & -65:35:16.1 & 310.307 & -3.755 & 0.21 & 11.57 & 10.91 &  10.82 &   13.87  &  12.99 &   12.77 &  2.88 & 3.45 & 0.40 &  4.64 &  0.0289  \\
 VVVX-J140431.82-655950.5 & 14:04:31.82 & -65:59:50.5 & 310.279 & -4.174 & 0.16 & 12.52 & 11.92 &  11.90 &   15.28  &  14.44 &   14.21 &  3.19 & 2.70 & 0.35 &  3.65 & 0.0361  \\
 VVVX-J140545.14-654103.8 & 14:05:45.14 & -65:41:03.8 & 310.488 & -3.909 & 0.17 & 10.27 & 9.54  &   9.49 &   12.44  &  11.62 &   11.45 &  3.15 & 3.83 & 0.36 &  4.53 &  0.0184  \\
 VVVX-J141036.24-653457.9 & 14:10:36.24 & -65:34:57.9 & 310.997 &  -3.958 & 0.19 & 9.96 &  9.25 &   9.18 &   12.94  &   12.06 &  11.85 &  5.46 & 3.72 & 0.38 &  4.36 &  0.0162 \\
 VVVX-J141605.01-651502.8 & 14:16:05.01 & -65:15:02.8 & 311.644 &  -3.821 & 0.24 & 10.47 &  9.97 & 10.06 &   13.27 &    12.25 & 12.06 &  3.58  & 3.73 & 0.35 & 6.21 &  0.0148 \\
 VVVX-J141829.93-643221.4 & 14:18:29.93 & -64:32:21.4 & 312.119 &  -3.232 & 0.29 & 11.80 & 11.36 & 11.42 &   14.68 &    13.71 &  13.62 & 3.33  & 3.16 & 0.29 & 6.29 &  0.0246 \\
 VVVX-J142155.59-643952.6 & 14:21:55.59 & -64:39:52.6 & 312.423 & -3.473 & 0.32 & 11.94 &  11.57 & 11.63  &   14.68 &    13.69 &  13.51 & 2.79  & 3.32 & 0.67 & 2.76 &  0.0247 \\
\hline
\hline
\end{tabular}
\label{tab:satellites}
\tablefoot{The identification is shown in column (1), the J2000 equatorial coordinates in columns (2) and (3), the Galactic coordinates in columns (4) and (5), the A$_{Ks}$ interstellar extinction in column (6), total extinction-corrected $J^{0}$, $H^{0}$, and $K_{s}^{0}$ total magnitudes in columns (7) to (9), the extinction-corrected $J_{2}^{0}$, $H_{2}^{0}$, and $K^0_{s2}$ aperture magnitudes within a fixed aperture of 2~arcsec diameter in columns (10) to (12), the morphological parameters $R_{1/2}$, $C$, ellipticity and $n$ in columns (13) to (16), respectively,  and the photometric redshift estimation in column (17).}
\end{sidewaystable*}

\begin{table*}
\caption{Absolute $K_s$ magnitudes and half-light radii statistics.}
\begin{tabular}{l|c|ccc|ccc}
\hline
\hline
Galaxies                  &  N  & \multicolumn{3}{|c|}{M$_{K_s}$[mag]} & \multicolumn{3}{|c}{R$_{1/2}$[arcsec]} \\
                          &     & median    & min     & max     & median & min & max\\
\hline
R$_{1/2}$ $>$ 2.45 arcsec & 20  &   -24.05 $\pm$ 0.59 & -25.22 & -23.09 &  3.00 $\pm$ 0.74 & 2.45 & 5.46 \\
R$_{1/2}$ $>$ 2.45 arcsec \& $z_{phot}$ $<$ 0.04  & 8        &   -23.90 $\pm$ 0.61 & -25.05 & -23.39 &   3.33 $\pm$ 0.84 & 2.79 & 5.46 \\
\hline
\end{tabular}
\label{tab:luminosityrhalf}
\end{table*}

\section{The distribution of galaxies around the Circinus galaxy}
    \label{section5}

        \begin{figure*}
    \centering
    \includegraphics[width=1\textwidth]{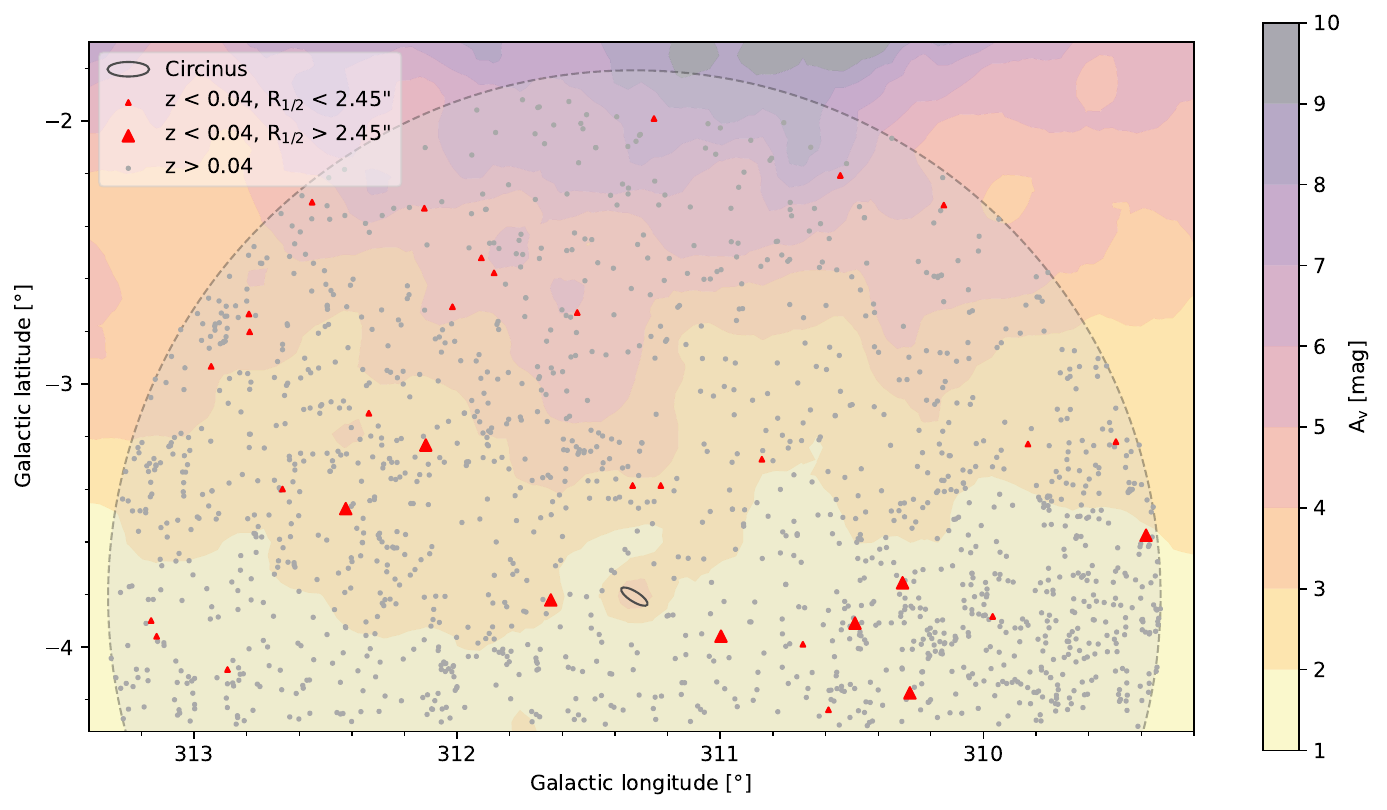}
    \caption{%
    The distribution of galaxies in Galactic coordinates around the Circinus galaxy. The  triangles represent galaxies with photometric redshifts lower than 0.04, and the size of the triangles is related to the half-light radius of the galaxies. The grey dots represent the galaxies with photometric redshifts greater than 0.04. }
    \label{fig:zphotfinal}
    \end{figure*}

The photometric redshift estimates for the VVV NIRGC galaxies in the studied region, together with their spatial distribution in the sky, allow us to search for potential spatial groupings or local overdensities of galaxies around the Circinus galaxy.
 
Figure~\ref{fig:zphotfinal} shows the spatial distribution in the sky of the 1,542 VVV NIRGC
galaxies in the studied region. The background is colour-coded by the A$_V$ interstellar extinction, and the Circinus galaxy is represented by an ellipse, as shown in Figure~\ref{fig:MAP1}. Galaxies with $z_{\mathrm{phot}} < 0.04$ are represented by red triangles of different sizes depending on the half-light radius. The eight of these galaxies  with R$_{1/2} > $ 2.45 arcsec  
are shown with the biggest triangles and they  appear scattered across the field, without any clear overdensity or clustering near the Circinus position. However, these sources are still significantly distant from the systemic redshift of the Circinus galaxy ($z \sim 0.0015$), and none of them can be confidently associated with its immediate halo. 

Galaxies with $z_{\mathrm{phot}} > 0.04$ are shown as grey circles in the figure. These remaining galaxies are consistent with being background sources, and are likely to be part of the large-scale structure located behind the Galactic plane. Their spatial distribution reflects both the intrinsic clustering of distant galaxies and the impact of spatially varying interstellar extinction, which affects the depth of detections across different regions of the field.
   
The VVV NIRGC galaxies reached 2 magnitudes deeper than the 2MASS Extended Source Catalogue (2MASSX; \citealt{Jarrett2004}), although we were unable to find Circinus satellites.  
In addition, the study of \cite{2019MNRAS.484.3334S} has revealed that the mid-infrared dust emission from Circinus AGN appears prominently extended in the polar direction. This result suggests that the polar dust wind is indeed optically thick, significantly contributing to the obscuration, which might explain the scarcity of satellites in the Circinus neighbourhood.


\section{Summary and final remarks}
\label{section6}

In this study, we searched for satellite galaxies within a circular area of a 2-degree radius around the Circinus galaxy, the nearest type-2 Seyfert galaxy. The radius of the studied region represents roughly 2/3 of the Circinus virial radius, and
a total of 1,542 galaxies were selected from the VVV NIRGC I and VVV NIRGC III catalogues, and visually inspected using $J$, $H$, and $K_{s}$ images of both the VVV and VVVX surveys. The limiting magnitude was $K_{s}$ = 15.5 mag, with a completeness of detections of about 90\%, 
corresponding to an absolute magnitude limit of $M_{K_s} = -12.6$ mag.
We identified 20 candidate satellites for the Circinus galaxy that have a half-light radius in the range 2.45 $ < R_{1/2} < $ 5.46 arcsec, consistent with expected angular sizes for dwarf satellites found in galaxies. 

We implemented, for the first time, the photometric
redshift estimates in this extinct region
of the sky due to the presence of the Galactic disc using machine learning techniques applied
to NIR photometry. The ANNz algorithm was
trained on a carefully selected subset of galaxies
with known spectroscopic redshifts and photometry of both VVV and VVVX surveys.  
This procedure enabled us to obtain
reliable photometric redshifts, particularly in the
range $0.001 < z_{phot} < 0.023$, where the calibration was most effective.
This range includes the Circinus galaxy ($z\sim 0.0015$), and we found eight low-redshift
galaxies ($z_{phot} < 0.04$) with sizes of typical dwarf satellites of $ R_{1/2} > $ 2.45 arcsec, 
but far from the Circinus galaxy. %

In the studied region, within a 2-degree radius around the Circinus galaxy,
a resolved
satellite system around the galaxy could not be
verified using the current data. However, this does not rule out the
existence of faint dwarf satellites. Given the large
angular size, high surface brightness, and complex
structure of the Circinus galaxy itself, it is
plausible that low-luminosity satellites remain undetected because of projection effects, high interstellar extinction, and detection difficulties due to the bright host galaxy.

The methodology  developed here constitutes a significant step 
forward.  
The estimated uncertainties for $z_{phot}$ of around $\sim 0.005$ ensured the effective removal of distant background galaxies.  
This technique provides a foundation for
more complete future studies of low-redshift 
galaxies behind the Galactic disc, and confirms that 
even challenging areas of the sky can be incorporated into extragalactic science.

In regions of low Galactic latitudes, photometric redshift techniques are essential because spectroscopic data are either severely limited or absent. While less precise than spectroscopic measurements, these estimates effectively exclude high-redshift galaxies, thereby refining candidate galaxy samples. This approach is the most realistic and effective technique to overcome the significant observational challenge in the zone of avoidance, and will enable us to expand our understanding of the nearby Universe. Spectroscopic follow-up observations are essential for confirming galaxies and conducting precise dynamical studies. When combined with the photometric methods presented herein, this will substantially enhance our ability to map the distribution of galaxies, in general, and also the faint satellite systems of nearby galaxies, in particular. This will improve our understanding of galaxy formation and evolution within the broader cosmological context.


\begin{acknowledgements}

The authors would like to thank F. Rodríguez, D. García Lambas, and M. Abadi for useful discussions that greatly improved this work. This study was partially supported by Consejo de Investigaciones Cient\'ificas y T\'ecnicas (CONICET) and Secretar\'ia de Ciencia y T\'ecnica de la Universidad Nacional de C\'ordoba (SeCyT).
D.M. gratefully acknowledges support from the Center
for Astrophysics and Associated Technologies CATA by the ANID BASAL
projects ACE210002 and FB210003, and by Fondecyt Project No. 1220724.
M.S. acknowledges support from ANID’s FONDECYT Regular grant No. 1251401. 
J.G.F-T gratefully acknowledges the grant support provided by ANID Fondecyt Iniciaci\'on No. 11220340, ANID Fondecyt Postdoc No. 3230001 (Sponsoring researcher), from two Joint Committee ESO-Government of Chile grants under the agreements 2021 ORP 023/2021 and 2023 ORP 062/2023.

We gratefully acknowledge the use of data from
the ESO Public Survey program IDs 179.B-2002 and 198.B2004 obtained
with the VISTA telescope, and optical data products from the Cambridge Astronomical
Survey Unit (CASU), the VISTA Science Archive (VSA), 
and the ESO Science Archive. VVV and VVVX data are published in
the ESO Science Archive in the data collections identified by the following
DOIs: https://doi.eso.org/10.18727/archive/67 and
https://doi.eso.org/10.18727/archive/68.
This research has made use of the SIMBAD database,
operated at CDS, Strasbourg, France.

\end{acknowledgements}

\bibliographystyle{aa} 
\bibliography{aa55715.bib} 

\end{document}